\documentclass[aps,prl,twocolumn,groupedaddress]{revtex4}
\begin{document}


\title{The simplest Bell's theorem, with or without locality} 


\author{Charles Tresser}
\email[]{charlestresser@yahoo.com}
\affiliation{IBM, P.O. Box 218, Yorktown Heights, NY 10598, U.S.A.}


\date{\today}

\begin{abstract}
We prove here a version of Bell's Theorem that is simpler than any previous one.  The contradiction of Bell's inequality with Quantum Mechanics in the new version is not cured by non-locality so that this version allows one to single out classical realism, and not locality, as the common source of all false inequalities of Bell's type.  
\end{abstract}

\pacs{03.65.Ta}

\maketitle

%
%
%
%
Consider a sequence of spin-$\frac{1}{2}$ particles, each prepared to have positive spin along the arbitrary vector $\vec{a}$ (hence normalized spin $s= +1$ along  $\vec{a}$); thus we have a constant sequence $\{s_i(\vec{a})=+1\}$. In order to fix notations, we recall that Quantum Mechanics (QM) teaches us that if one measures the spin of these particles along vector $\vec{b}$, one gets a sequence $s'_i(\vec{b})$ such that the average value $<s_i(\vec{a})\cdot s'_i(\vec{b})>$ (which is also  $\langle  s_i(\vec{a}) |  s'_i(\vec{b})\rangle$ in Dirac's notations) is equal to $\cos (\vec{a},\,\vec{b})$.  This is the spin-$\frac{1}{2}$ version of Malus' law for polarization, which may be better known in terms of probabilities of coincidences:  
\[
\pi (s_i(\vec{a})= s'_i(\vec{b}))=\frac{1+ <s_i(\vec{a})\cdot s'_i(\vec{b})> }{2}= \cos ^2 (\frac{(\vec{a},\,\vec{b})}{2})\,.
\]

Following  Bohm's  treatment of the EPR Paradox \cite{EPR}, \cite{Bohm}, we deal with \emph{electron-positron pairs} and study the \emph{spins} of the particles using \emph{measurement tools} (MTs). On each MT, one can choose a   \emph{MT vector} so that the spin of a spin-$\frac{1}{2}$ particle is either along or opposite to the MT vector, with normalized projection $s$ equal to either +1 or -1 (with probability one, a precision that is elsewhere implicit whenever it applies).  \emph{The pairs are prepared, in the  \emph{singlet state}  (\cite{Bohm}, p. 400) whose spin part is rotation invariant and given along any vector by
$$
\Psi(x_1,x_2)=\frac{1}{\sqrt{2}}(| +\rangle _1\otimes| -\rangle_2-| -\rangle_1\otimes|
+\rangle_2)\,,
$$
so that for any direction, the total spin is 0}.  There is a similar setting for light polarization that is more practical for experiments but that will not be used here \cite{BohmAharonov1957}.  

The electrons are observed using the MT $E$, and the positrons using the MT $P$,  these MTs being operated by people or smart machines. Fix $\vec{x}$ and let $X_i(\vec{x})\in\{\mathcal E_i, \mathcal P_i\}$ be the normalized spin measured by $X\in \{E,P\}$ along $\vec{x}$ on the corresponding $i^{\rm{th}}$ element of a sequence of pairs in the singlet state. Thus the total spin zero property reads: 
\[ 
\forall \vec{x},\, \forall \vec{i}, \quad E_i(\vec{x})+P_i(\vec{x})=\mathcal E_i+ \mathcal P_i=0\,.
\]

\medskip
In any Bell's type argument we assume:

\noindent 
- i) Either the strong hypothesis of Bell in \cite{Bell} (see also \cite{Bell2004}) that there are Hidden Variables (HVs) such that the statistical properties are the same as for usual QM, and which are \emph{predictive} (\emph{i.e.,} the future values of observables are determined by the present state of the world if the present state is described using all the needed variables, including the HVs, even if no one can predict nor access all of these values).   

\noindent
- ii) Or a weaker hypothesis, obviously implied by the former one but not equivalent to it: \emph{Classical realism} such that the statistical properties are the same as for usual QM and whereby an observable has a value that is well defined whenever one is assuming that the measurement would be made, even if the measurement is eventually not performed (for instance because another measurement is performed instead).  

A \emph{Bell's Inequality} (BI) is an inequality between some number of objects such as 
$<X_i(\vec{x})\cdot Y_i(\vec{y})>$ or such as $\pi(X_i(\vec{x})= Y_i(\vec{y}))$ (or of the form $\pi(X_i(\vec{x})= Y_i(\vec{y})=1)$ that is not used in this paper), where $X,\,Y\in\{E,\, P\}$ while $\vec{x}$ and $\vec{y}$ belong to a finite collection of names of vectors, collection whose size depends on the particular BI version.  The vectors' names correspond to fixed or variable vectors, depending on the particular BI derivation.  For any derivation of a BI,  we will adopt the convention that there are two vectors associated to the same symbol, say $\vec{a}$, if the vector $\vec{a}$ is a MT vector of both $E$ and $P$.  We will characterize the \emph{complexity of a BI} as \emph{the number of vectors being used in its derivation, with the stated convention}.  I propose here the first derivation of a BI using only 3 vectors.  

A \emph{Bell's Theorem} (BT) is (depending on the authors) either the contradiction obtained by replacing the objects such as $<X_i(\vec{x})\cdot Y_i(\vec{y})>$ or $\pi(X_i(\vec{x})= Y_i(\vec{y}))$ in a BI by their values provided by QM (for some special choices of the vectors if the vectors were arbitrary in the BI), and/or the implications of such a contradiction. Thus a BT deals with potential extensions or modifications of QM, some of which get disqualified as a mean to avoid the contradiction in the (form of) BT being considered.

Such contradictions usually disappear, as do the formerly  known BIs themselves, if one drops the assumption of \emph{locality} whose effect is that \emph{the values of the observables near one MT are not influenced by the settings of the other MT}.  The general meaning of locality is that effects cannot propagate faster than light: the setting independence that has been stated then follows whenever the evaluations of observables are spatially separated.  The new BI derivation proposed here resists dropping the locality assumption, so that we will challenge the usual claim that BTs indicate non-locality: see \cite{Tresser2} and \cite{Tresser} for related material.

\medskip
I now recall one of the former simplest BI derivations (using 4 fixed vectors): it has been adapted from \cite{Stapp1971} and told by Mermin to Penrose \cite{Penrose1989}, \cite{Penrose2004}.  The presentation is made in two parts since the first part only will be used below in the discussion of the new BT.

\medskip
- I) Following \cite{Penrose1989}  and \cite{Penrose2004}, suppose that there are:

- 2 MT vectors for $E$, $|\rightarrow~\rangle$ at  $0^\circ$, and $|\uparrow \rangle$ at  $90^\circ$ (we use kets for vectors: the context should tell if a given ket represents a vector, a spin state along that vector, or possibly both entities),

- 2 MT vectors for  $P$, $|\nearrow \rangle$ at $45^\circ$ and $|\searrow \rangle$ at $- 45^\circ$.

Take the \emph{actual} settings to be respectively $|\rightarrow \rangle$ for $E$ and $|\nearrow \rangle$ for $P$ for a first run of experiments, and call $\mathcal E=\mathcal E_1,\, \mathcal E_2, \dots$ and $\mathcal P=\mathcal P_1,\, \mathcal P_2, \dots$ the runs respectively registered by $E$ and $P$.  Then QM tells us that the probability $\pi(\mathcal E_i = \mathcal P_i)$ that $\mathcal E_i$ and $\mathcal P_i$ coincide is 
$
\frac{1}{2}(1-\cos (45^\circ ))=0.146...\,,
$
hence \emph{about} $15\%$ (in the sequel, I drop the word ``about" in front of percentages or probabilities since the sizes of the corresponding approximations and errors are much smaller than what would be needed to reverse the strong inequalities that show up in the discussion).  The MQ prediction comes from \emph{Wave Packet Reduction} according to which the entangled singlet state becomes the product state 
$\Psi'(x_1,x_2)=|+_{\vec{a}} \rangle _1\otimes| -_{\vec{a}}\rangle_2$ or $\Psi'(x_1,x_2)=|-_{\vec{a}} \rangle _1\otimes| +_{\vec{a}}\rangle_2$ after the $E$ measurement along $\vec{a}=|\rightarrow \rangle$ is performed. Then $P_i(\vec{a})=-E_i(\vec{a})$ and the conclusion follows from Malus' law.

\medskip
Next we assume \emph{locality} so that $\mathcal P$  does not depend on the $E$ setting.  Now call $\mathcal E'$ the run that would have been registered by $E$ if the alternate MT vector $|\uparrow \rangle$ had been chosen (under assumption i) or ii)).  The percentage $\pi(\mathcal E'_i =\mathcal P_i)$ of agreement between $\mathcal E'$ and $\mathcal P$ would then have been again equal to
$
 \frac{1}{2}(1-\cos (45^\circ ))=0.146...\,.
$

\medskip
- II) The story continues as follows: if the $E$ settings had been $|\rightarrow \rangle$  as initially decided, but the $P$ settings had been $|\searrow \rangle$, then the run at $E$ would have been $\mathcal E$ as before, using again locality.  Denoting by  $\mathcal P'$ the runs that would have been registered by $P$ with the new setting of its MT vector  (under assumption i) or ii)), the expectation of coincidence $\pi(\mathcal E _i=\mathcal P'_i)$ between $\mathcal E_i$ and  $\mathcal P'_i$ would have been again
$
\frac{1}{2}(1-\cos (45^\circ ))=0.146...\,.
$
The punch line will then come from the difference between two ways of comparing the runs $\mathcal E'$ and  $\mathcal P'$:  

- On the one hand the agreement of these runs could not be better than
$45\%=15\%+15\%+15\%$ as a result of the above discussion (hint: all
entries being binary, for $\mathcal E'_i$ and $\mathcal P'_i$ to
agree, one needs at least one agreement between $\mathcal E'_i$ and
$\mathcal P_i$, between $\mathcal P_i$ and $\mathcal E_i$, or between
$\mathcal E_i$ and $\mathcal P_i'$, whence the $45\%$ bound): this $45\%$ bound  
is the Bell's Inequality that we were after.  

- On the other hand a direct computation according to QM yields:
$
\pi(\mathcal E'_i = \mathcal P'_i)=
\frac{1}{2}(1-\cos (135^\circ ))=0.854...\,,
$
a clear contradiction with the $45\%$ bound that is the Bell's Theorem's contradiction that was aimed at. 

- The Bell's Theorem's conclusion then states that \emph{one at least of classical realism or the locality assumption is false}, with locality in the role of the usual suspect (as clearly stated by Penrose in \cite{Penrose2004}, p. 589, following a tradition going back to \cite{Bell}).  The fact is that \emph{without locality, $\pi(\mathcal E'_i = \mathcal P'_i)$ makes no more sense and the conclusion vanishes}, but this hardly proves that locality is the essential hypothesis. 

\medskip
\emph{The counter-natural character of the reasoning implies that no conclusion of a BT can be verified experimentally, to the contrary of the main claims in \cite{Aspect1999} and references therein.}  By a \emph{counter-natural} I mean a \emph{gedanken} experiment such that some law of physics would need to be violated to permit it (this is related to but different from counterfactual); the adjective is defined accordingly.  

\smallskip
Furthermore, \emph{only realism or HVs that are too naive to be compatible with QM have been disqualified by BTs or by the related arguments.} This partial but essential emptiness of  the original BT was pointed out as far back as 1972 in \cite{PenaCettoBrody1972}, a paper that is summarized on page 312 of Max Jammer's  treatise \cite{Jammer1974}. The analysis in \cite{PenaCettoBrody1972} relates to the argument in \cite{Bell} and attacks the lack of sensitivity, to former measurements, of the measure density that allows one  to take averages such as correlations; Bell's response to \cite{PenaCettoBrody1972} in \cite{Bell1975} falls short of answering the critic.  The parallel to the analysis in \cite{PenaCettoBrody1972} for proofs assuming ii) consists in noticing that the form of classical realism being used is incompatible with QM \cite{Tresser2} (see also \cite{Tresser}). 

\smallskip
Too bad for Bell's type theorems, but I nevertheless propose below a new proof that is simpler than the previously simplest one reported above since it uses only part I of that proof.  Simple versions of (\emph{gedanken}) experiments can help us better understand what is going on in a given problem in physics: for an analysis of the essential equivalence of all Bell's type theorems, see  \cite{Fine1982a} and \cite{Fine1982b}.  The probabilistic approach of Fine was later continued by Pitowsky (see \cite{Pitowsky1994}, \cite{Pitowsky2001} and references therein).  

\medskip
\emph{The new argument only uses part I)  of the presentation made above of the simplest formerly known version of BT}.  In particular, \emph{we assume locality until otherwise stated.}  

\smallskip
\noindent
{\bf No Correlation Lemma.} \emph{The sequences $\mathcal E$  and $\mathcal E'$ are not correlated,} i.e., : 
\[
(\circ)\qquad <\mathcal E_i \cdot \mathcal E'_i >=0 \quad \Leftrightarrow\quad\pi(\mathcal E_i = \mathcal E'_i)=\frac{1}{2}\,.
\] 

\noindent
\emph{Proof of the No Correlation Lemma (local case):} Under the locality assumption, $(\circ)$ was already proved by Stapp (see his formula 7 in \cite{Stapp1971}), and would have readily followed from the arguments of Bell in \cite {Bell}. Just notice that with $\mathcal E''$ standing for what would be observed along $|\uparrow \rangle$ on the $P$ side,  by QM we have $ \pi(\mathcal E_i = \mathcal E''_i)=\frac{1}{2}$. Conservation of the spin and augmenting QM by HVs or classical realism then yields $\mathcal E'=-\mathcal E''$, from where $(\circ)$ follows readily. This proves the lemma under the locality assumption. 

As Arthur Fine pointed out to me, such a derivation not only uses but also requires locality. I provide below a proof of the No Correlation Lemma with no locality assumption, since such a proof is needed to establish the Main Claim.  Meanwhile I proceed with my BT assuming locality.

- A)  \emph{Data collection:} we consider the triple of pairs $(\mathcal P,\,\mathcal E)$,  $(\mathcal P,\,\mathcal E')$, and  $(\mathcal E,\,\mathcal E')$. For each of the two first pairs, we had previously obtained a $15\%$ chance of coincidence, while the chance of coincidence is $50\%$ for the third pair by the No Correlation Lemma.

- B)  \emph{BI version 1:} One way of reasoning is to notice that $\mathcal P_i$ coincides with one of $\mathcal E_i$  and $\mathcal E'_i$ whenever $\mathcal E_i \not = \mathcal E'_i$.  Otherwise speaking:
$$
(*)\,\, \min [ \max (\pi(\mathcal P_i = \mathcal E_i),\, \pi(\mathcal P_i = \mathcal E'_i))]\geq \frac{1- \pi(\mathcal E'_i = \mathcal E_i)}{2}\,.
$$

- C)  \emph{BI version 2:} Alternatively, we can notice that the sum of the three percentages of coincidences must be over $100\%$ 
, or 
$$
(**)\qquad \pi(\mathcal P_i = \mathcal E_i)+\pi(\mathcal E_i = \mathcal E'_i)+\pi(\mathcal E'_i = \mathcal P_i)\geq 1\,.
$$
for any of $\mathcal P$, $\mathcal E$ and $\mathcal E'$ to coincide with itself (by the same argument that has generated a $45\%$ bound for $\pi (\mathcal E'_i=\mathcal P'_i)$). Formula $(**)$ is formula 2a of \cite{Fine1982b} (and formula 4 of \cite{Pitowsky1994}) specialized to probability 0.5 for any measurement of a single spin $\frac{1}{2}$.  

-  \emph{B \& C's conclusion (from BI to BT):} Whichever BI, $(*)$ or $(**)$, one uses to show that \emph{the $15\%$-$15\%$-$50\%$ triplet of percentages of coincidences provides a false inequality}, this proof of a BT is the simplest that can be by the number of MT vectors that it requires (as for $(*)$, notice that its right hand side is at least $25\%$ by $(\circ)$).    

- D) \emph{Further comments:} Notice that $(**)\Rightarrow (*)$.  Furthermore, by using the rotational invariance of the singlet state, one readily generalize $(*)$ to $pa(\theta)\geq \frac{1}{2}(1-pp'(2\theta))$  where $pa(\omega)$ is the probability of having the same reading for the two particles being measured along two MT vectors that are $\omega$ apart and $pp'(\omega)$ is the probability of getting twice the same values for one of the particles along two MT vectors that are $\omega$ apart: for $\theta=45^\circ$, one gets back the contradiction derived above. In a similar way, $(**)$ gets generalized to $pa(\theta)+pa(\theta')+pp'(\theta+\theta')\geq 1$, which is still a special case of formula 2a in \cite{Fine1982b} and for which we just set $\theta=\theta'=45^\circ$ to get back the contradiction associated to the $15\%$-$15\%$-$50\%$ triplet.

\medskip
For any choice of the details (B, C, or D), the (setting of the) proof that has just been presented here is simple enough to easily let appear the counter-natural character of the reasoning.  Unlike what happened with all former settings of BTs we have the following: 

\smallskip
\noindent
\textbf{Main Claim:} \emph{Non-locality would be of no help in the proposed setting 
and only the counter-natural character of the \emph{gedanken} experiment can be the cause of the problem}. 

\smallskip
\noindent
\textbf{Corollary to the Main Claim:} Thus \emph{it is the counter-natural character of the reasoning, as permitted by the strong Bell hypothesis or by classical realism, that is the \textbf{only} problem common to \textbf{all} the versions of Bell's Theorem including the one presented in this paper.} 

\smallskip
\noindent
\emph{Proof of the Main Claim:} Assuming that non-locality effects can change correlations to help suppress the impossible triplets of coincidences:

- For a Galilean observer who first sees the $P$ side where one measures or just uses i) or ii):

$\quad$ - a1) The probability $\pi (\mathcal P_i=\mathcal E_i)$ is $15\%$ from QM,

$\quad$ - b1) The probability $\pi (\mathcal P_i=\mathcal E'_i)$ is $15\%$ from QM,

$\quad$ - c1) The probability $\pi (\mathcal E_i=\mathcal E'_i)$ would need to be at least $70\%$ from a1) and b1) in view of $(**)$, but we have no independent way to know.

- For a Galilean observer who first sees the $E$ side where one measures or or just uses i) or ii):

$\quad$ - a2) The probability $\pi (\mathcal E_i=\mathcal E'_i)$ is $50\%$ by the No Correlation Lemma, which is  proved in the non-local case below. 

$\quad$ - b2) The sum $\pi (\mathcal P_i=\mathcal E_i)+\pi (\mathcal P_i=\mathcal E'_i)$ would need to be at least $50\%$ from a2) in view of $(**)$, but we have no independent way to know.

\noindent
When the $E$ and $P$ sides are considered synchronously in the Galilean frame of the experiment (well defined although possibly no measurement is performed), one then gets what one is sure to get in each of the previously mentioned asynchronous cases (because the outcomes cannot change according to the Galilean frame even if non-locality cannot be formulated in a Lorentz-invariant way).  Thus in the synchronous case one keeps (only) the firm conclusions   a1), b1), and a2) from the two asynchronous cases so that one gets back the same $15\%$-$15\%$-$50\%$ triplet that was found to hold true and to cause a contradiction with a BI when locality was assumed.  This concludes the proof of the Main Claim.

\smallskip
\noindent
\emph{Proof of the No Correlation Lemma (non-local case):} 

- $(\alpha )$ If none of the vectors $|\rightarrow \rangle$ and $|\uparrow \rangle$ is used on the $E$ side to perform an actual measurement, we introduce the further vector $|-\uparrow \rangle$, to which would correspond the sequence $\mathcal E''=-\mathcal E'$. This implies that $\pi(\mathcal E_i = \mathcal E'_i)+\pi(\mathcal E_i = \mathcal E''_i)=1$ so that if for instance $\pi(\mathcal E_i = \mathcal E'_i)\geq\frac{1}{2}$, one has $\pi(\mathcal E_i = \mathcal E''_i)\leq\frac{1}{2}$.  We use here sequences that are unknown, but known to be well defined by (naive) classical realism or HVs.  The only thing that could matter and generate inequalities is the orientations of the angles $(|\rightarrow \rangle, |\uparrow \rangle )$ and $(|\rightarrow \rangle, -|\uparrow \rangle)$ .  Since the  vectors $-|\uparrow \rangle$ and $|\rightarrow\rangle$ play the same role as long as no measurement is performed, only equality can happen.  More precisely, using $-|\uparrow \rangle$ as the primary vector instead of $|\rightarrow \rangle$, one would find $|\rightarrow \rangle$ in the role payed so far by $|\uparrow \rangle$ from which the equality follows immediately (assuming isotropy of physics).

- $(\beta )$ If to the contrary one does measure along one of the vectors, say $|\rightarrow \rangle$, we notice that one arrives at the same conclusion $(\circ)$ by using invariance under parity and the fact used in $(\alpha )$ that only the orientation of the angle could matter. 

This concludes the proof in the most important non-local case, and we notice that the measurements need not be made so that the use of parity invariance can be avoided if one prefers.  Charge conjugation invariance would also lead to $(\circ)$ by showing the equivalence of the two angle orientations, but by using both the $E$ and the $P$ sides, which cannot help in the non-local case.

\smallskip
\noindent
{\bf Remark.} \emph{We have used a trick when dealing with the synchronous measurements (or inquiries) on the $E$ and $P$ sides, but that trick does not work in the example recalled above from Penrose's books.  The difference is that all the pairs that are considered in the new proof come about even if without locality, while this does not apply to the pair $(\mathcal E', \,\mathcal P')$ of the former simplest example, since $\mathcal E'$  comes about in conjunction with $\mathcal P$, while $\mathcal P'$  comes about in conjunction with $\mathcal E$. Only locality can make the pair $(\mathcal E', \,\mathcal P')$ take life if  $(\mathcal E, \,\mathcal P)$ represents what has been actually performed in terms of measurements (starting with another couple would only displace the problem).}

\medskip
\emph{It follows from Bell's type results that to be acceptable in view of QM, realism or HVs theories need to be less naive than the form used here (often wrongly attributed to Einstein because of \cite{EPR}); with the new proof, one needs no more to believe in locality to reach such conclusions.}  One can check that when Einstein personally used realism after 1935 in texts designed to be made public, he used it in a (non-naive) way that respected the fact that one cannot access conjugate variables on a particle.  This is in contrast with the text of the EPR paper \cite{EPR} which is recognized (see \cite{Jammer1974}, \cite{FineShaky}) to have been written by Podolsky and disliked by Einstein who did not even see the last version before it got published.  The EPR paper is very different from Einstein's own expositions of the problem of the completeness of the wave function as a descriptor of states (the problem that is the subject of \cite{EPR}).  These expositions of the completeness problem avoid counter-naturals: see, \emph{e.g.,} the 1936 text reproduced in \cite{EinsteinIdeasAndOpinions} and the much later text pp. 83-87 in \cite{Schilpp} where Einstein stays clear from any classical realism trap.  As for HVs, Rosen stated in 1985 (see \cite{Rosen1985}) that HVs were never part of the picture in the minds of the authors of \cite{EPR}, and Einstein stated about the HVs theories of de Broglie and of Bohm that such approaches were too naive \cite{BornEinstein}: see also \cite{Tresser} and \cite{Tresser2} where these issues are discussed with much more details, and \cite{Jammer1974} and \cite{FineShaky} as historical sources.  
\begin{acknowledgments}
\textbf{Acknowledgments:} 
The support of my family, friends, nurses and doctors was so essential that I cannot thank them enough.
I also thank Michel le Bellac, Arthur Fine, Bob Griffiths, Larry Horwitz, Christian Miniatura, Itamar Pitowsky, Itamar Procaccia, Oded Regev, Edward Spiegel, and  Lev Vaidman  for reading or listening to versions of this work and suggesting many words and improvements.
\end{acknowledgments}

%
%

\begin{thebibliography}{99}
\bibitem{EPR}
A.\ Einstein, B.\ Podolsky, N.\ Rosen \textit{Phys. Rev.} \textbf{47}, 777 (1935).

\bibitem{Bohm}
D.\ Bohm \textit{Quantum Theory,} ( Prentice Hall; New York 1951).

\bibitem{BohmAharonov1957}
D.\ Bohm, Y.\ Aharonov \textit{Phys. Rev.} \textbf{108}, 1070 (1957).

\bibitem{Bell}
J.S.\ Bell, \textit{Physic} (Long Island City, NY)  \textbf{1}, 195 (1964).

\bibitem{Bell2004}
J.S.\ Bell, \textit{Speakable and Unspeakable in Quantum
Mechanics,}  (Cambridge University Press; Cambridge, Revised
Edition 2004).

\bibitem{Tresser2}
C.\ Tresser \textit{Any is not all: EPR and the Einstein-Tolman-Podolsky paper},  quant-ph/0503006, to appear.

\bibitem{Tresser}
C.\ Tresser \textit{Weak realism, counterfactuals, and decay of geometry at small scales}, quant-ph/0502007, to appear.

\bibitem{Stapp1971} 
H.P.\ Stapp, \textit{Phys. Rev.} \textbf{3 D}, 1303 (1971).

 \bibitem{Penrose1989}
R.\ Penrose, \textit{The Emperor's New Mind,} (Oxford
University Press, Oxford; 1989/ \ Books, New York; 1991).

\bibitem{Penrose2004}
R.\ Penrose, \textit{The Road to Reality,} (Jonathan Cape, London; 2004).

\bibitem{Aspect1999}
A.\ Aspect, \textit{Nature} \textbf{398}, 18, March (1999).

\bibitem{PenaCettoBrody1972}
L.\ de la Pe{\~n}a, A.M.\ Cetto, T.A.\ Brody,  \textit{Lett. Nuovo Cimento} \textbf{5} 177 (1972).	

\bibitem{Jammer1974}
M.\ Jammer, \textit{The Philosophy of quantum mechanics: The
Interpretations of quantum mechanics in Historical Perspective,}
(John Wiley \& Sons Inc; New York, 1974).

\bibitem{Bell1975}
J.S.\ Bell, \textit{Epistemological Letters,}  Nov. 1975, 2-6; see also \cite{Bell2004}, pp.63-66.

\bibitem{Fine1982a}
A.\ Fine, \textit{Phys. Rev. Lett.} \textbf{48}, 291 (1982).

\bibitem{Fine1982b}
A.\ Fine, \textit{J. Math. Phys.} \textbf{23}, 1306 (1982).

\bibitem{Pitowsky1994}
I.\ Pitowsky, \textit{Brit. J. Phil. Soc.} \textbf{45}, 95 (1994).

\bibitem{Pitowsky2001}
I.\ Pitowsky, \textit{Phys. Rev.  A} \textbf{64}, 4102 (2001).

\bibitem{FineShaky}
A.\ Fine, \textit{The Shaky Game; Einstein Realism and the Quantum Theory,}
(The University of Chicago Press; Chicago, 2$^{\rm nd}$ edition, 1996).

\bibitem{EinsteinIdeasAndOpinions}
A.\ Einstein, \textit{Ideas and Opinions,} (Crown Publishers; New York, 1954).

\bibitem{Schilpp}
P.A.\ Schilpp, (editor) \textit{Albert Einstein:
Philosopher-Scientist} (The Open Court Publishing Co.; La Salle,
IL, 3$^{\rm rd}$ edition, 1969 (Vol. 1) - 1970 (Vol. 2)).

\bibitem{Rosen1985}
N.\ Rosen, in \textit{Symposium on the Foundations of Modern Physics: 50 years of the Einstein-Podolsky-Rosen Gedankenexperiment,}  P. Lahti and P. Mittelstaedt, Eds.  (World Scientific; Singapore, 1985), pp. 17Ð33. 

\bibitem{BornEinstein}
M.\ Born, A.\ Einstein, \textit{The Born-Einstein Letters; Correspondence between Albert Einstein and Max and Hedwig Born from 1916 to 1955 with commentaries by Max Born,} I. Born Transl., (Walker and Company; New York, 1971).
\end{thebibliography}
%
%

\end{document}